# NEURAL COMPUTATION AT THE THERMAL LIMIT

Levy, W.B.[1,*], Berger T.[2], and Fleidervish, I.A.[3]

[1] Department of Neurosurgery, University of Virginia, Charlottesville, VA, 22908
[2] Department of Electrical and Computer Engineering, University of Virginia, Charlottesville, VA, 22904
[3] Department of Physiology and Cell Biology, Faculty of Health Sciences and Zlotowski Center for Neuroscience, Ben-Gurion University of the Negev, Beer Sheva, 84105, Israel

[*] Communicating author email: wbl@virginia.edu; or postal delivery: Dept. of Neurosurgery, Health Sciences Center, Univ. of Virginia, Charlottesville, VA, 22908.

ABSTRACT

Although several measurements and analyses support the idea that the brain is energy-optimized, there is one disturbing, contradictory observation: In theory, computation limited by thermal noise can occur as cheaply as ~$2.9 \cdot 10^{-21}$ joules per bit (kTln2). Unfortunately, for a neuron the ostensible discrepancy from this minimum is startling – ignoring inhibition the discrepancy is $10^7$ times this amount and taking inhibition into account >$10^9$. Here we point out that what has been defined as neural computation is actually a combination of computation and neural communication: the communication costs, transmission from each excitatory postsynaptic activation to the S4-gating-charges of the fast Na+ channels of the initial segment (fNa's), dominate the joule-costs. Making this distinction between communication to the initial segment and computation at the initial segment (i.e., adding up of the activated fNa's) implies that the average synaptic event reaching the fNa's is the size of the standard deviation of the thermal noise, $(kT)^{1/2}$. Moreover, defining computation as the addition of activated fNa's yields a biophysically plausible mechanism for approaching the desired minimum. This mechanism, requiring something like the electrical engineer's equalizer (not much more than the action potential generating conductances), only operates near threshold. This active filter modifies the last few synaptic excitations, providing barely enough energy to allow the last sub-threshold gating charge to transport. That is, the last, threshold-achieving S4-subunit activation requires an energy that matches the information being provided by the last few synaptic events, a ratio that is near kTln2 joules per bit.

*Author's significance statement*
In what sense does the brain compute? Given the optimizing nature of natural selection and 0.6 billion years of neural evolution, one expects that microscopic neural processes are optimal. The extent optimizations support minimizing energy costs of communication and computation. However, under pervasive, simplifying assumptions, a neuron, described by the signal transformation it performs, seems far from optimal, >$10^8$ times the limit set by statistical thermodynamics. Carefully distinguishing communication costs within a neuron from computational costs produces a new definition of microscopic neuronal computation and reveals that communication accounts for most all the extremely large costs (99.998%). Via this reanalysis, one aspect of computation is predicted to occur near the best possible joules per bit limit set by physics.

\body
INTRODUCTION
One path for understanding and interpreting neural computation consists of identifying optimizations that Nature (evolution through natural selection) has achieved. In general organisms are not optimal, but bits and pieces of organisms can be on average optimal when a local environment (e.g., the internal, physiological milieu of the organism) has been unchanged for eons. Thus it is natural to presume [e.g., 1-10] that the costs of computation and communication, at the level of single neurons and networks that are generic across mammalian species, are energy-optimized. In this regard there are a few first principle successes [e.g., 7,8,11,12], but there is one glaring failure facing the energy-efficient hypothesis.

Unfortunately for the energy-efficient approach, a nominally critical calculation fails[4,13,14]. Specifically, calculations from physics and from Shannon theory place the optimal energy efficiency of computation at *kTln*2 ≈ 2.9·$10^{-21}$ J/bits (with *k* Boltzmann's constant and *T* the absolute temperature of a mammal, *ca*. 310° kelvin). Thus, the whole energy-efficiency program for interpreting brain computation can be called into doubt, perhaps replaced by arguments favoring a combination of compromised optimizations, including energy-efficiency [e.g., 14]. However here, after recapitulating the fundamental nature of the apparent failure of the hypothesis, we distinguish aspects of neural processing that imply a particular aspect of single neuron computation is near the thermal limit.

To simplify the presentation and to highlight the most important assumptions and hypotheses, a number of relevant but distracting observations and relationships are in the Supplement [Sa-Sh]. The subsections to follow will (A) define a neuron, (B) explain the limit kTln2 J/bit, (C) display the gross inefficiency of neural computation via a typical analysis and then pointing out a particular, minor inefficiency of an additive computation, (D) discover an aspect of synaptic excitation that is near the noise level, (E) carefully distinguish communication from computation (F) uncover another aspect at the thermal limit via interpreting molecular events as computation, i.e., activation of the fast Na-channels of the initial segment and transport of their gating charges. Figs 1 and 2 provide a visual summary of the beginning and ending models.

DEVELOPMENTS, COMPARISIONS, AND A NOTABLE MOLECULAR HYPOTHESIS

(A) *A simple neuron defined*

At the level of a single, simplified neuron (Fig 1a), neural computation is (1) the combining of input states by something akin to addition (or, indeed, addition itself), then (2) producing a pulse-out when threshold is reached, and (3) reinitializing the membrane potential. To this time honored definition of a McCulloch and Pitts (McC-P) neuron[15], we add the assumption of interpulse interval (IPI) coding of a neuron's latent variable[16,17,Sa]. But this will not be enough for our analysis because the physical limit on computation arises from certain practical considerations not part of the McC-P neuron. In particular, the neuron of interest computes (adds and resets) with charge accumulation and dissipation for which there are resistive and capacitive paths between the neuron's inside and outside. Finally in order to approach observed anatomy and physiology, our neuron (Fig 1b) only makes sense when it also has inhibitory inputs and leak[18,Sb].

(B) *A brief review of kTln2 J/b and its relevance*

The kTln2 limit on dissipative sensor-memory systems arises with Szilard but his argument is now replaced by a memory storage perspective of Landauer and of Bennett as described in [19] (see also a recent unification[20]).

Because of a neuron's RC nature (R:=resistance, C:=capacitance) and from the fact that its repetitive use requires excitatory charging followed by erasure with each output pulse, a neuron is dissipative and is thus subject to the *kTln2* limit. Specifically, (1) a neuron starts an IPI at its reset/resting potential; (2) it senses each of its active excitatory synaptic events as positive charge injections; (3) then it stores these charges on its membrane capacitor while at nearly the same time (4) subjecting these positive injections to scaled inhibitory shunts to ground; and with enough such charge accumulation, (5) a pulse-out is generated; at which point, (6) the neuron's potential returns to rest. This last step is an energy dissipating step in which the capacitive charge associated with $Na^+$ fluxes is countered by a $K^+$ conductance and its associated current flow. Thus a neuron's computation is energy dissipating. Moreover, the reason it takes at least *kTln2* joules to load and unload one bit of computational information can be related to the resolvable stored-signal steps where the resolution of these steps can be no better than what is allowed by random capacitive charging driven by the ambient, Gaussian thermal noise, i.e., a zero-mean noise whose variance is directly proportional to *kT* .

Like the physics result but from a very different perspective, electrical engineers also arrive at this same relationship (but inverted) growing out of the classic communication problem [21(fig 1)] for channel capacity. Starting with Shannon's theorem 17 [22] for the capacity of a channel with bandwidth B Hz and average signal power, $P_s$, capacity$=B\log_2(P_s/P_n+1)=B\ln(P_s/(BkT)+1)/\ln(2)$ bits/s, where noise-power, $P_n$, is thermal noise per Hz so that the ratio inside the logarithm is unitless. Dividing by the average signal power $P_s$ per second and taking the limit of $\ln(1+P_s/BkT)$ (either the small signal limit or the large B limit) yields the desired relationship

$$\lim_{P_s \to 0} \frac{capacity}{P_s} = B\frac{P_s/BkT}{P_s \cdot \ln 2} = \frac{1}{kT\ln 2} \; bits/J \;\; \text{where the units evolved from} \; \frac{bits/s}{J/s}.$$

In summary, the two perspectives lead to the same limit that interrelates information and energy, and the relevance of this limit to a neuron is established through the neuron's RC character.

(C) *A simple example of the costly nature of neural computation*

To illustrate the existence of a disturbingly large joule expense of neural computation when using the simple neuron, suppose $V_m$, the voltage of our simple neuron at its initial segment, has to be moved 16 m$V$ to fire from the reset $V_m$ (e.g., reset produces the resting potential of –66m$V$ and an output pulse occurs at the time-instant when $V_m$ first attains –50m$V$). The energy to reset each time is, crudely, the energy to charge the membrane capacitor, $C_m$, from reset to threshold: $C_m V^2/2 = 3 \cdot 10^{-10}$ $F$ $(.066^2 - .050^2)$ $V^2 \approx 2.7 \cdot 10^{-13}$ $J$ per output pulse, where we have used 300 p$F$ for $C_m$.[24] Throughout assume 6 $b$/pulse (8 Hz average firing rate arising from a *ca.* 0.125 ms temporal resolution; cf. [17,23]), implying (J/pulse divided by $b$/pulse) $4.5 \cdot 10^{-14}$ J/$b$. Thus, at minimum, the expenditure is 17 million times more costly than the so-called minimum of $kT\ln 2$, *ca.* $2.9 \cdot 10^{-21}$ J/$b$, a disconcerting excess under the energy-efficient hypothesis. (In fact, having ignored inhibition and leak, this estimate is ca. 100-fold too low).[Sb, Sc]

In general, models of neural computation rely on the central limit theorem (CLT) to generate information[e.g.,7] via additivity and approximate conditional independence. Unfortunately generating information in this fashion is costly although no where near as costly as calculated above. Specifically, via the CLT, information grows with the number of active synapses, $n$, as $2^{-1}\log n$ while energy costs grow in proportion to $n$. Thus, when there are 5000 inputs lines (see below), the cost of computation will differ from $kT$ by at best, ca. 100-fold [Sd] whenever all the activated synapses are considered. However, certain aspects of individual synaptic events, and particularly those events near the end of the IPI may come quite close to $kT$.

(D) *Identifying a place where kTln2 is approached*

Our first approach to $kT$ requires two disparate calculations: the size of the average synaptic event arriving at the initial segment vs. the thermally induced noise voltage. Synaptic events at the thermal level are revealed. This leads us to carefully distinguish communication from computation and then to quantify the two processes responsible for the >$10^7$ discrepancy.

Instead of starting with the synaptically localized charge influxes, consider the following line of reasoning that takes the forebrain cortical physiological situation into account, i.e., the case of a functioning neuron under heavy synaptic bombardment that includes excitation and inhibition. Such heavy synaptic bombardment must, on average, be the normal state of affairs for a neuron of the neocortex. How much excitation is there? Quite a lot as the first of the two calculations shows.

Suppose every excitatory neuron receives $n=10^4$ excitatory input lines and that each of these input neurons fires at 10 Hz, i.e., <u>once every 0.1 s</u> on average, and so does any one of their postsynaptic target neurons. The amount of inhibition is not specified as yet but it is sufficient to yield a relationship that does not contradict the premises just given. With a 10 Hz average firing rate at the output, the average IPI of this neuron is also 0.1 s; therefore, there will be — on average — $10^4$ excitatory input activations occurring at a single neuron over the IPI of this output neuron because each of the *n* inputs fires, on average, exactly once in the 0.1 second IPI of the postsynaptic neuron. Indeed, if there are *n* inputs and the average firing rate is any number you choose for both the input neurons and the output neuron, there will be, on average, *n* activations per IPI so long as the input and output firing rates are the same (as is reasonable for a neuron in a recurrent network). In what follows we will use *n*=5000 which seems reasonable for a rat neocortical neuron as does 8 Hz as an average firing rate. Now for the second of the two disparate calculations.

Because *kT*, the background thermal noise-energy per two degrees of freedom, has little intuitive meaning from the viewpoint of $V_m$, a simple conversion of these joules to volts will be both helpful. This conversion hinges on a very gratifying relationship. Although the noise-energy, proportional to *kT*, is generated by the resistive components of a neuron and varies with resistance, it is the noise-voltage across the membrane capacitor that we care about in terms of how this noise limits information-processing by a neuron. From this perspective, $V_m$ as a voltage across the membrane capacitor, the relationship is quite simple: The mean-square thermal voltage-fluctuations across the plasma membrane capacitor, $C_m$, do not depend on any of the values of the resistive paths to ground; the thermally induced variance of $V_m$ is just $kT/C_m$ $V^2$ (see, e.g. [25], the calculation boils down to the autocorrelation function[Se] at zero delay and applies to any passive circuit regardless of its complexity). Thus one can ignore the membrane resistance, which certainly is not constant and ignore any bandwidth calculation, which is already incorporated into this variance. Again taking the value of $C_m$ as 300 p*F*, the thermally induced variance of the neuron's voltage is $kT/C_m \approx 4.2 \cdot 10^{-21}$J$/300 \cdot 10^{-12}F = 1.4 \cdot 10^{-11}$ $V^2$, and thus its standard deviation is, $\sqrt{14 \cdot 10^{-12} \ V^2} \approx 3.8 \cdot 10^{-6} \ V$.

We now combine the distinct calculations comparing the average synaptic event from the perspective of $V_m$ relative to the noise-voltage; remarkably the two are now quite similar. Thus, to detect one bit of synaptic information, we require an average synaptic event to inject enough charge to depolarize $V_m$ about one or two standard deviations of the thermally induced noise; that is, we compare the value of 3.8 μ*V* to the size of the average excitatory synaptic contribution invested specifically into moving the $V_m$ to threshold. Crucially, this voltage contribution is not calculated from the charge injection at the synapse *per se*. Instead, the calculation uses the average number of excitatory synaptic events and uses the voltage difference between reset and threshold. Specifically, the 5000 on-average excitatory synaptic events move $V_m$ from rest to *16mV* depolarized from the resting potential. Thus, and under the contentious but simplifying assumption of linear summation, to achieve threshold the average synaptic event contributing to this sum moves the membrane potential $16mV \div 5000 = 3.2 \cdot 10^{-6} V$ (or $6.4 \cdot 10^{-6} V$ with a 50% failure rate). Therefore, with these assumed values, the average synaptic event that adds to the $V_m$ of the initial segment is just about $\sqrt{kT/C_m}$. (Perhaps this result is 'too good to be true', and certainly some might argue for $n$=10,000 while another might argue for 2500 input lines, but these other values just lead to adjustments that are quite small compared to the original comparison which misses by a factor of >10$^7$.)

Thus we have a calculation that reveals, in one sense, how computation does occur in the vicinity of the thermal limit. Before moving onto to the second sense in which one finds a computational operation at $kT$, we reconcile this first perspective with the high cost of neural computing as conventionally viewed. Along the way, we make explicit the argument that the greatest non-axonal cost, like the axonal cost itself [26,27] is interpretable as a communication cost.

(E) *The core hypothesis: distinguishing synapto-initial segment communication from initial segment computation*

It is well understood that a neuron's dendrite has a large, apparently unavoidable surface area [13] leading to large membrane capacitance, and in turn, leading to large energy costs. We can combine this empirical knowledge with the computational hypotheses developed here.

The hypothesis, or viewpoint of this section seems a natural extension of the previous calculation: Instead of defining neural computation as the addition of synaptic events from the perspective of the synapses, dichotomize this single perspective into communication and computation. Before computation begins, a large fraction of the synaptic energy is for communication. Before assessing the energy and information of such distinctions, let us make clear the first part of the dichotomy, communication, by interpreting the postsynaptic communication from a Shannon perspective.

Messages are created with each arriving input-pulse. The postsynaptic message is $\{\tau_{n_i}, t_{n_i}, w_{ij}\}$ where $\tau_{n_i}$ is the arrival time of the $n_i$ synaptic activation along input-line $i$, $t_{n_i}$ is the signaled IPI, and $w_{ij}$ is the conductance magnitude of the associated synaptic event (proportional to the charge injection). This message (1) is encoded as a voltage-transient, (2) is transmitted through a channel that consists of the dendrite and soma, losing energy, adding noise, and distorting the voltage-transient, which transient (3) reaches the initial segment to alter $V_m$ and increases the probability of activating fast Na$^+$-channels (fNa's). A literal decoder at the initial segment need not exist; nevertheless one can speculate about the existence of a filter (e.g., a voltage-dependent equalizer) that recovers, from the distorted synaptic message, much of the original information.

The relative energy loss due to $C_m$, i.e., communication, is a function of the total available, functional capacitance ($C_{IS}$) due to the fNa's, i.e., $C_{IS} \div C_m$.[sf] However because $C_{IS}$ is a functional capacitor, it only exists when charge transport occurs. To be precise, such transport occurs over a voltage range, nevertheless, the error of approximating this voltage range as a point seems small. Assume that gating-charge transport occurs on average around -52mV when the range of averaging ends at the assumed threshold of -50mv (at which voltage 50% of the 5.2e per fNa[28] have transported and threshold is achieved). Recalling: capacitance is just the ratio q/V; that there are $1.6 \cdot 10^{-19}$ coulombs/$e$; and the 750 fNa's at the initial segment, $C_{IS}=3.1 \cdot 10^{-16}C/0.052V=6.0 \cdot 10^{-15}F$ where $3/4 \cdot 10^3 \cdot 1/2 \cdot 5.2e \cdot 1.6 \cdot 10^{-19}C/e=3.1 \cdot 10^{-16}$ C. Returning to the ratio of interest, $C_{IS} \div C_m = 2 \cdot 10^{-5}$, and this ratio is not affected by shunting inhibition, which will, however, multiply energy costs. That is, 99.998% of the energy from synaptic activations can be viewed as the cost of communicating synaptic events to the initial segment, and this fraction is true regardless of how much additional energy arises through considerations of inhibition and leak, whose smaller cost multiple we associate with the inefficiency of information produced through the CLT. Moreover, even considering the S4 fNa's activated at threshold (two per fNa[28]) still puts us far from kTln2 ($.052V \cdot 3.1 \cdot 10^{-16}$ $C \div 6b \approx 2.7 \cdot 10^{-18}$ J/$b$). However, consider the last of the molecular events.

(F) *A molecular hypothesis*

Although there is no denying that both adding charge to the membrane capacitor and using leak/shunting inhibition as divisor of this sum is computation (as noted in Fig. 2), let us now take the extreme viewpoint that the essence of a neuron's computation is the successive addition of activated fNa's or activation, individually, of energy-barrier-crossing events of single fNa's (e.g., an S4 subunit or even, individually, each positive charge of an S4 crossing the focused[29] voltage-field). Whatever the case, such activations are summed until a threshold number of activated fNa's is reached. That is, in such extreme versions of the computational hypothesis, computation begins – and ends – at the fNa's of the initial segment: charge injections provided by equilibration with the changing values of $V_m$ are summed over the fNa's gating-charge transport until threshold fNa activation is achieved (and then the available, remaining fNa's activate). (Curiously with this conjecture, all aspects of the interpreted computation are essentially digital.)

Suppose, with some simplification, that $V_m$ is near but just one fNa-activation below threshold, and suppose many of the fNa's are but one, nominal S4 activation away from turning on. Then consider the energy involved in switching a single such S4 subunit from its partially depolarized "off-state" to its fully "on-state." Let this operation require a gating-charge[28] of 2.6$e$ (=$4.2 \cdot 10^{-19}$ coulombs) to cross $V_m$ at threshold (ca. -50 m$V$). Such charge transport requires 50 mV $\cdot 4.2 \cdot 10^{-19}$ coulombs =$2.1 \cdot 10^{-20}$ joules; in other words, 4.9$kT$ or 5.6 *kTln2* J. As our goal is to reveal some computational process that produces *kTln2* J/$b$, this last S4 energy requirement leaves us with the question: Under what model is the energy invested in this last switching event worth just a little less than the bit-value of the entire IPI itself?

At the outset we stated that excitatory synaptic responses are combined together with something like addition and then treated all responses as being the same and added together. This is consistent with the hypothesis that the neuron is estimating the value of its latent variable averaged uniformly across the IPI, but it is inconsistent with the assumed RC characterization of the neuron. Therefore, consider another model:

Suppose the neuron is performing the filtering problem of control theory; that is, the neuron is trying to estimate, with minimum mean square error, a nonstationary latent variable's value at the end of the IPI[30,31]. In this case the excitation of the neuron should be exponentially discounted over time. From a biophysical perspective, this discounting can correspond to the RC, low-pass nature of the neuron. With such exponential discounting, the early excitatory synaptic events will contribute little to reaching threshold; temporal weighting (via intrinsic and inhibitory shunting) emphasizes the late events. Indeed along with this exponential discounting, there are other biophysical effects that further emphasize the last few synaptic events; for example, the slowly inactivating Na-channels of the soma and apical dendrite [32] will, due to their voltage-dependence, increasingly amplify synaptic events as threshold is approached.

Combining this alternate functional model with the qualitative statements of neural discounting and amplification, we conjecture that just the last few (*ca*. five) synaptic excitations are required to cross with a precision (temporal and amplitude-scaling) for producing 6 *b/pulse* (a bit-match to the energetic requirement of 6 *kT*ln2 J of the conceptualized last, and threshold achieving, S4 activation event). 'Temporal and excitation precision' refers to a recovery of arrival times (up to a fixed delay) and amplitude (up to a scalar constant) of these last few synaptic messages $\{\tau_{n_i}, t_{n_i}, w_{ij}\}$ by something like an equalizer[Sg]. Sufficient precision, *ca*. 125 μsec as a guess, imbues the time of threshold-crossing with the required amount of information: Recall that (1) the time to threshold-hitting/crossing contains all the information of the IPI; (2) each input line provides information at 6 *b/pulse*, so there is plenty of extra information available in five events (on average); and (3) as a set, these five events occur in about 125 μsec (on average, as before, 5000 events are spread out over 125 msec). In sum, the conjectured model has most of its 6 bits generated by the most accurate, last S4 event, closely matching the 5.6·*kT*ln2 J needed for this last event to yield the desired ratio *kT*ln2 J.

Thus we have presented two computational viewpoints and several hypotheses consistent with neural computation near the thermal limit.

DISCUSSION

The separation, for the purpose of optimization[6,7], of communication and computation at the level of the axon versus all the rest of the neuron seems without controversy. Here we start the analysis by assuming the axon can be optimized independently of the rest of the neuron[6] (implicitly then, requiring the computational portions of the neuron to produce information at this axonal optimal rate) and then go on to consider the nature of computation and communication for what remains, all in the context of minimizing J/bit. Separating communication from computation for the rest of the neuron (synapses, dendrites, soma, and initial segment) seems to have been less than obvious until one adds a modicum of physical realism to the McC-P neuron. The added complexity includes: (*i*) an IPI coding for the output; incorporating (*ii*) noise-producing resistances as well as (*iii*) capacitive paths to ground; and eventually (*iv*), the fNa's – all of which provide a critical context for the developments here.

As a contribution, the above presentation includes, in addition to the proposal for a minimal upgrading of the McC-P neuron, two or three quantifications. Furthermore, there is an unintended result arising from the large communication costs; this quantification clearly emphasizes the importance of research concerned with optimizing a dendritic tree[12,33-35]. But the intended outcome of the research here is also fulfilled.

One now can counter the indictment that Nature is suboptimal when it comes to J/bit invested into computation *per se*. As a result, it is sensible to continue a research program that interprets neural computation through such optimizations[e.g., 1,8, 16,36]. Importantly, we propose interpretations of neural function consistent with neural computation occurring near the *kT* limit, particularly when taking the perspective of just the fNa's at the initial segment and the precision of threshold crossing. In addition, the analysis points out the cost of computation that generates information[Sd] through central-limit-theorem controlled convergence. Finally, by interpreting neuronal function through the assumption of energy-efficient computation, an apparently novel interpretation of neuronal computation results[Sh]. Thus, there is justification for further studies of optimized computation, even studies that ignore the cost of dendritic communication.

Nevertheless, in spite all the useful outcomes and apparently novel perspective, we still hear:
 "What a disappointing result. Why should anyone care about energy-efficient computation when 99.998% of the synaptic energy is going to communication?" While this might be the right question for a practical engineer, this is, of course, the wrong question for a biologist. The pertinent question a biologist asks is "Why should Nature care about energy-efficient computation when 99.998% of the energy is going to communication?" And asking the question this way leads to a more-than-plausible speculation.

Simply put, do not be misled (as we initially were) by the neocortex of rats and humans. One must go back quite far, much more than 200 Myr, to obtain the appropriate perspective. We hypothesize that, at early, seminal times of animal evolution, long and extensively branched dendrites and especially their great cost due to surface area are non-existent; relatively speaking, such dendrites are a recent innovation compared to somato-initial segment computation. In evolutionary history, we have no idea how few synapses the first spike-generating, decision-making neuron had, but surely that number was small – who knows: Was it 1,2, or 20 independent synapses? Thus the conjecture is that Nature evolved energy-efficient computation because synapto-initial segment communication was a modest fraction of the non-axonal total cost 600 Myr's ago. That is, computation was optimized before dendritic communication grew to become an equal and then the dominating, non-axonal energy cost. This conjecture is consistent with the general philosophy: Nature gets things right microscopically and then builds upon such success without losing the earlier design successes.

Acknowledgements: Costa Colbert and Rob Baxter generously critiqued many versions of this manuscript and its slowly evolving ideas. The research was partially funded by the NSF and by the Dept. of Neurosurgery, UVa.

Figure Legends.

Fig 1. Comparing the original computational model neuron to the ultimate neuron of this paper.
(A) For the Mc-P neuron all the communications are axonal. Synaptic responses over the IPI are added, $\sum_i \sum_{n_i} w_{ij} 1_{\tau_{n_i}}(\tau)$, and as this sum accumulates, this sum eventually reaches threshold at which point a pulse is generated and the neuron is reset. (See Fig 2 legend for notation). (B) The final model subdivides postsynaptic function. Although the functions may overlap in space and share microscopic biophysical processes, the functions are partitioned into 1) synaptic activations, 2) communication of these activations to the initial segment, 3) production of $V_m(\tau)$, which includes shunting inhibition, and 4) gating-charge movements under the influence of the communicated and transformed synaptic events, which leads to fNa activations that, in turn, lead to output pulse production and reset. More details are found in Fig 2. (Thermal noise is neglected in this figure but not in the model).

Fig 2. More details of the new neuron of Fig 1B and some notation used in the text. Call the neuron of interest *j*; its task is to estimate the value $\lambda$ of the latent random variable $\Lambda$. Each neuron has its own perspective on time. For *j*, denote its "running" time with the variable $\tau \in [0, t_j]$; that is, $\tau = 0$ corresponds to the beginning of a new IPI and $\tau = t_j$ corresponds to the end of this IPI. During this IPI, neuron *j* receives pulses from its inputs, indexed by *i*; implicitly, each arriving pulse implies an IPI $t_{n_i}$. Moreover, associated with each arrival is (1) the synaptic strength $w_{ij}$ and (2) the time of arrival $\tau_{n_i}$ where the double subscript is necessary to allow for multiple pulse arrivals on input line *i* during *j*'s current IPI. Functionally $V_m(\tau)$ can be described multiple ways: 1) an exponential moving average (and thus both a) computation and b) memory storage); 2) an implicit decoder (i.e., the pair $\{V_m(\tau), \tau\}$ can be used to estimate of $\lambda$, $\hat{\lambda} = V_m(t_j) \div t_j$; and 3) as a voltage-dependent equalizer at the end of the IPI.

Fig 1
### A. McC-P neuron

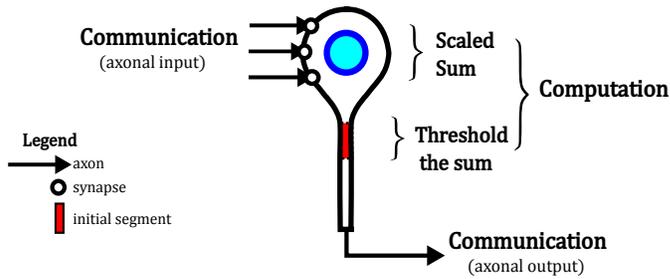

### B. Proposed neuron

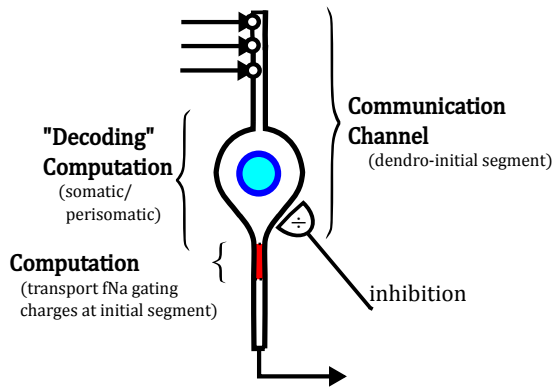

Fig 2

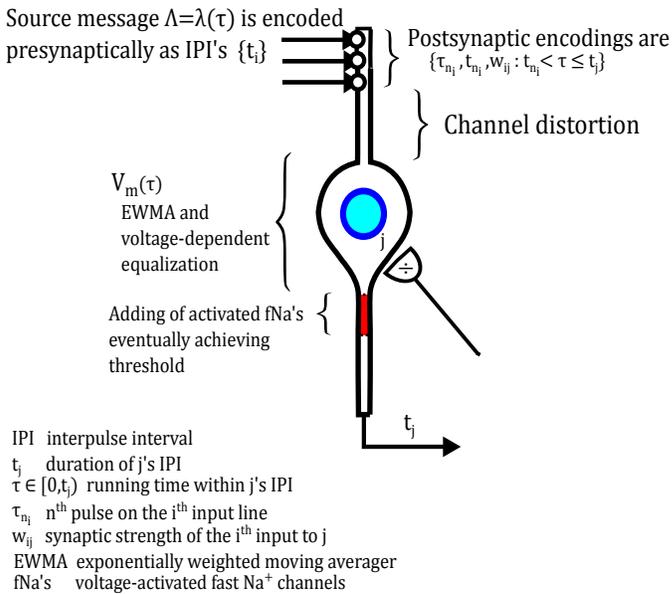

IPI  interpulse interval
$t_j$   duration of j's IPI
$\tau \in [0,t_j)$  running time within j's IPI
$\tau_{n_i}$  $n^{th}$ pulse on the $i^{th}$ input line
$w_{ij}$  synaptic strength of the $i^{th}$ input to j
EWMA  exponentially weighted moving averager
fNa's  voltage-activated fast $Na^+$ channels

SI Text

[a]*The latent variable*
We assume that a neuron is computing a positive scalar latent variable. Specifically, the latent variable controls the rate of build-up of $V_m$ through the variable's influence on the rate of firing of the excitatory inputs. The latent variable is unobservable and only inferable through $V_m(s)$, $s \in (0,t.)$. For the simple case that a neuron $j$ adds its inputs, the best estimate of $\lambda_j$ at the end of the IPI is $\hat{\lambda} := \theta \div t$. Thus a recipient of $t$ with knowledge of the threshold $\theta$ has information equivalent to this value's best estimate. Moreover, given an appropriate probability model with $t$ a sufficient statistic and $\theta$ an ancillary statistic, then $t$ contains all the information of the IPI. [S1]

By virtue of its selected inputs ($i \in \{1,2,...,n\}$) and their synaptic weights ($w_{ij}$), neuron $j$ is constructed to estimate the value of a particular latent variable for each of its output IPIs. The value $\lambda_j$ of random latent variable $\Lambda_j \in (0,\infty)$ is implicit in the mean value of the point processes of the $n$ inputs to $j$ during the IPI. (For each $i$, the pulse-arrival-times on each input line are realized random variables arising from the associated point process). Most simply is the case of stationary point processes with mean values $\gamma_{ij}$, $\lambda_j := \sum_{i=1}^{n} w_{ij} \gamma_{ij}$. In the case of point processes that are nonstationary over the IPI [0,$t$), $\lambda_j(t) := t^{-1} \int_{s=0}^{t} \sum_{i=1}^{n} w_{ij} \gamma_{ij}(s) ds$ is sometimes used. A variety of nonstationary models need to be considered starting with $\forall i, d\gamma_i(s) = (\mu(s) - \gamma_i(s))ds + \sigma dW_i$ where the $W_i$ are independent Weiner processes, $\mu(s)$ is a random or deterministic drift, and $\sigma > 0$ scales the Wiener process.

[b]*Inhibition and leak raise costs*
For excitatory synaptic excitation to produce the energy equivalent of 16 mV of depolarization on the membrane capacitor, a large amount of energy must be dedicated to offsetting inhibition and leak. Suppose that the average synaptic event is 160pS. Further suppose that a rested neuron under zero inhibition requires 4 nS of excitatory synaptic activation to reach threshold, i.e., 25 synaptic activations. Suppose a neocortical failure rate of 50%. Then of the, on average, 5000 active inputs, 2500 will be active. Therefore inhibitory shunting needs to downgrade this number of activations 100-fold to reach the equivalent of the 25 activations when there is no inhibition.

[c]*The relevant energy costs*
As we are trying to assess the cost of raw computation, many other, necessary costs are ignored. All of our calculations explicitly exclude presynaptic costs including transmitter release (replenishment, pumping, etc) and the cost of spike generation once threshold is obtained, both of which we group with axonal costs. Maintenance, repair, growth, and all other forms of overhead are also ignored. For a full assessment of energy costs [see S2].

*dInformation production allowed by inhibition and leak*

In the sense of improving the accuracy by which a neuron estimates its latent variable, $\hat{\lambda} = \theta \div t_j$ (where $t_j$ is the IPI and $\theta$ is the threshold value of $V_m$), inhibition and leak enable greater sampling and thus greater accuracy (the standard deviation is growing as the square root of $t$). The rate at which information increases with $n$ (the number of active synapses) is essentially proportional $2^{-1} \log_2(n)$; thus the necessary 6 bits would arise when $n$ is ca. 4096 ($2^{-1} \log_2(2^{12})$).

*eAssumption for $C_m$ noise calculation*

There is one assumption in this calculation that can only be approximated. The autocorrelation (covariance) is properly obtained by integrating over all time, i.e., from minus to plus infinity. The value of this integral will be suitably approximated when a neuron has an appropriate "time constant" relative to an IPI, i.e., the constant governing the exponential rate of decay is something like 3-5 times shorter than the IPI. Note that the final result of this calculation is consistent with, for some, the initiutive guess $E[V_m^2]C_m = kT$.

*fDerivation of $C_{IS} \div C_m$*

As an approximation of the biophysics, consider the initial segment functional capacitance and the cell membrane capacitance to be charged to the same steady-state voltage from the same resting voltage. Then $\dfrac{C_{IS}}{C_m} = \dfrac{C_{IS} \cdot \int_{v_{rest}}^{v_{final}} v\, dv}{C_m \int_{v_{rest}}^{v_{final}} v\, dv}$.

*gThe Hodgkin-Huxley mechanism interpreted as an equalizer*

The conjectured equalizer is no more than the voltage-activated channels of the initial segment, cell body, and nearby main dendrite. These channels reshape synaptic responses when polarization is in the vicinity of threshold. Supporting the equalizer concept is the following observation. If an electrical engineer starts with an RGC transmission line (e.g., a passive axon-like submarine cable) and then adds a series of homogeneously spaced equalizers with enough gain to recover a pulse's amplitude, then the engineer has built a transmission line that is equivalent to an axon. That is, the line shows no decrement and constant shape of a conducted pulse just like a Hodgkin-Huxley axon that uses voltage-activated channels. Therefore, one description of the function of these channels is equalization.

*hNovel interpretation of neural computation*

Defining computation as the addition of activated fNa's seems to provide a multiplicative interpretation needed by Bayesian theories of neural computation [e.g., S3, S4]. Consider the Boltzmann-form governing a simple model of individual fNa activation. Suppose every synapse has a weight-dependent, constant shunt effect $w_{ij}Q_{ij}^-$ [see S3] and, when activated ($x_i$=1 else $x_i$=0), a charge injection, $w_{ij}Q_{ij}^+$. Then taking the Boltzmann form of the probable state (open divided by closed) and using a constant to convert from charge to voltage, one gets a multiplicative form that can serve as an odds-ratio calculation. For example,

$c_0 \prod_i \exp(-c_1 w_{ij}(x_i Q_{ij}^+ - Q_{ij}^-)) = c_0 \exp(-\sum_i c_1 w_{ij}(x_i Q_{ij}^+ - Q_{ij}^-))$. Clearly such a conjecture needs significant amounts of development.

Supplement References

S1. Levy, W. B., & Berger, T. (2012, July). Design principles and specifications for neural-like computation under constraints on information preservation and energy costs as analyzed with statistical theory. In *Information Theory Proceedings (ISIT), 2012 IEEE International Symposium on* (pp. 2969-2972). IEEE.

S2. Attwell, D., & Laughlin, S. B. (2001). An energy budget for signaling in the grey matter of the brain. *J Cereb Blood Flow & Metabol 21*(10): 1133-1145.

S3. Levy, W. B., Colbert, C. M., & Desmond, N. L. (1990). Elemental adaptive processes of neurons and synapses: a statistical/computational perspective, in *Neuroscience and connectionist theory*. Erlbaum, Hillsdale, NJ.

S4. Deneve, S., & Pouget, A. (2004). Bayesian multisensory integration and cross-modal spatial links. *J Physiol-Paris*, *98*(1): 249-258.